\documentclass[reprint, superscriptaddress]{revtex4-1}
\usepackage[T1]{fontenc}
\usepackage[english]{babel}
\usepackage{amsmath,amsfonts,amsthm}
\usepackage{graphicx}
\usepackage{booktabs}
\usepackage{xcolor}
\usepackage{pgfplots}
\usepackage{anyfontsize}
\usepackage{calc}
\usepackage{subfiles}
\usepackage{tikz}
\usepackage{braket}
\usepackage{lipsum}
\usepackage{wrapfig}
\usepackage[margin=0.75in]{geometry}
\usepackage{pdfpages}
\usepackage{fancyhdr}
	
\usepackage{chngcntr}
 
\counterwithout{figure}{section}
\counterwithout{table}{section}

\makeatletter
\AtBeginDocument{\let\LS@rot\@undefined}
\makeatother
\pgfplotsset{compat=newest}
\usetikzlibrary{arrows}
\usetikzlibrary{calc}

\begin{document}

\title{ 2-designs and Redundant Syndrome Extraction for Quantum Error Correction }
\author{Vickram N. Premakumar}
\address{Physics Department, University of Wisconsin-Madison, 1150 Univ. Ave., Madison, WI, USA}
\author{Hele Sha}
\address{Physics Department, University of Wisconsin-Madison, 1150 Univ. Ave., Madison, WI, USA}
\author{Daniel Crow}
\address{Physics Department, University of Wisconsin-Madison, 1150 Univ. Ave., Madison, WI, USA}
\author{Eric Bach}
\address{Computer Science Department, University of Wisconsin-Madison, Madison, WI, USA}
\author{Robert Joynt}
\address{Physics Department, University of Wisconsin-Madison, 1150 Univ. Ave., Madison, WI, USA}
\address{
Kavli Institute for Theoretical Sciences, University of Chinese Academy of Sciences, Beijing 100190, China}
\date{{\normalsize \today}}

\begin{abstract}
Imperfect measurement can degrade a quantum error correction scheme.  A solution that restores fault tolerance is to add redundancy to the process of syndrome extraction.  In this work, we show how to optimize this process for an arbitrary ratio of data qubit error probability to measurement error probability.  The key is to design the measurements so that syndromes that correspond to different errors are separated by the maximum distance in the signal space, in close analogy to classical error correction codes.  We find that the mathematical theory of 2-designs, appropriately modified, is the right tool for this.  Analytical and simulation results for the bit-flip code, the 5-qubit code, and the Steane code are presented. The results show that design-based redundancy protocols show improvement in both cost and performance relative to conventional fault-tolerant error-correction schemes in situations, quite important in practice, where measure errors are common. In the near term, the construction of a fault-tolerant logical qubit with a small number of noisy physical qubits will benefit from targeted redundancy in syndrome extraction.       
\end{abstract}

\maketitle

\section{Introduction}
A working quantum computer must be fault-tolerant: the reliability of all components must be considered, and appropriate measures taken for compensation of malfunctions \cite{Shor1996, Preskill1999}.  Error correction forms the core of this process.  Error syndromes are extracted and used to generate the information needed to set the computer back on the right path.
The syndrome information may itself contain errors because of imperfect measurements.  This aspect of fault tolerance was recognized at an early stage of the development of the theory. In the two most prominent syndrome extraction protocols, the remedy was to repeat measurements already made \cite{Steane1996, Divincenzo1996}, thereby building in redundancy.  This is a straightforward way to make the syndrome information more reliable.

In the last few years, more sophisticated schemes have been proposed.  Fujiwara \textit{et al.} noted that measuring more than a minimal set of stabilizers could be a more efficient way to extract reliable information and noted a possible connection to designs \cite{Fujiwara2014, Fujiwara2015}.  Ashikmin \textit{et al.}   generalized existence theorems and the quantum Singleton bound of standard quantum error correction (QEC) theory so that they take into account redundancy in measurements \cite{Ashikhmin2014}.  Crow \textit{et al.}  used redundancy in syndrome extraction to give thresholds for qubit performance in a coherent error-correction scheme \cite{Crow2016}.  

In this letter, we show how to optimize measurement redundancy by applying the theory of 2-designs,  \footnote{  2-designs are also known as balanced incomplete block designs.  Our sense of the word 2-design is not related to quantum 2-designs, which are probability distributions over quantum states.}, This is called design-based redundancy (DBR).  2-designs have long been used in classical error correction \cite{pless}. We investigate both a minimal redundant extraction scheme (MR) as well as the more comprehensive DBR.  

It is sufficient to use a simple model for the faults.  The key feature of this model is to separately define the qubit error probability $p_q$ and the syndrome measurement error probability $p_m$.  Different physical implementations of quantum computation will have very different values of the ratio $p_q/p_m$ and this will strongly influence the optimal DBR protocol.  In this work we will only consider the simple situation of a quantum memory that is periodically refreshed by error correction; more complicated scenarios with active gates would complicate the analysis but not introduce significant new concepts.  The discussion is restricted to stabilizer codes \cite{Gottesman1996}.  It may be possible to use 2-designs to improve other codes, but it appears to be more complicated.  We stress that DBR is equally applicable to measurement-based error correction and coherent error correction.  For definiteness, we will use the language of measurement-based QEC in this work.     

\section{Bit Flip Code}
This section is included in order to introduce the basic ideas of DBR.  It treats the elementary example of the 3-qubit bit flip code \cite{NielsenChuang}.  The three physical qubits store a single bit of quantum information and there is a probability $p_q$ of a bit flipping.  No other qubit errors are allowed.  Stabilizers $S_1 = Z_1 Z_2$, $S_2 = Z_2 Z_3$ and $S_3 = Z_3 Z_1$ can each be measured, always yielding $\pm 1$.  An incorrect measurement result is obtained with probability $p_m$.  The starting state (chosen arbitrarily in the code subspace) is $ |000 \rangle $; after a certain time 1 or more bits may flip and we measure a set of stabilizers, perhaps repeatedly.  We define an event $e$ to be the final state of the qubits together with the measurement results.  Each event $e$ has a probability $P(e)$ with $0 \leq P(e) \leq 1$ and a success factor $s(e) = 0$ or $s(e) = 1$ when the event is respectively uncorrectable or correctable.  For example, if the textbook procedure of measuring only the generators of the stabilizer group $S_1$ and $S_2$ is used, a possible event is $e_0= \{|001 \rangle ,+1,-1\}$.  This can be made fault-tolerant by repeating the measurements and using majority rules on the measurement results.  Assuming independence, the probability of this event is $P(e_0) = p_q (1-p_m)^2 (1-p_q)^2$  since there is 1 qubit error and 0 measurement errors.  $s(e_0) = 1$ since the information obtained from the measurements allows us to correct the error.  The total failure probability of an error correction protocol, including possible DBR, is $F = 1 - \Sigma_e P(e) s(e)$.  In addition, we define the cost C of a protocol to be the expected total number of stabilizer measurements in a correction cycle. 

The DBR protocol differs from both the simple protocol and its fault-tolerant extension in that one measures \textit{ the complete set of stabilizer group generators} $S_1$, $S_2$ and $S_3$.  This already builds in redundancy; if there is one measurement error then exactly one of the $S_i$ is equal to $-1$.  If 2 of the $S_i$ is equal $-1$ then there is a unique instruction as to which bit to flip back. (Here and henceforth we use $S_i$ both for the operators and for the result of measuring the operators.) The key point is that there is a unique signal even if there is a measurement error.  For this toy code, MR and DBR are identical; for large codes this is not the case.

Once protocols are established, then it is straightforward to sum over the events and compute the failure rates and costs.  In Table \ref{Table1} we tabulate the results to quadratic order in $p_q$ and $p_m$ and the cost to linear order.

\begin{table} 
	\begin{tabular}{ccc}
    \toprule
	Protocol & Failure Rate & Cost \\
	\midrule
	Minimal QEC & $2p_m - p_m^2 + 3p_q^2$  &  $2$ \\
	Fault-tolerant QEC  & $6 p_m^2 + 3 p_q^2$  &  $4 + 2 p_m$  \\
	DBR  &  $3p_m^2 + 3 p_q^2 + 9 p_m p_q$  &  3\\
	\bottomrule
	\end{tabular}
	\caption {Failure rate and cost for 3 error-correction protocols 
	for the 3-qubit bit-flip code.}
	\label{Table1}
\end{table}

The fact that minimal QEC has a linear term in $p_m$ is the signature that it is not fault-tolerant and is therefore not a candidate for a working computer.  More importantly, the fault-tolerant version of conventional QEC is always more costly than DBR and the failure rate for conventional QEC exceeds that of DBR whenever $6p_m^2 + 3 p_q^2 > 3 p_m^2 + 3 p_q^2 + 9 p_m p_q$.  This reduces to $p_m > p_q /3$, which is likely to happen in many implementations.  It is interesting that DBR is superior even for $p_m = p_q$, a case often considered.  

\section{2-designs}

The crucial feature of DBR for the bit flip code is that the syndrome possesses a unique signature when there is a measurement error.  This feature can be generalized to more complex codes that can correct phase flip as well as bit flip errors.  Consider an $[[n,k,d]]$ stabilizer code with $n$ the number of physical qubits, $k$ the number of logical qubits and $d$ the distance; the code can correct errors on up to $(d-1)/2$ physical qubits.  We will focus on $k=1$ and the logical operators are $X_L=\otimes_{i=1}^{n} X_i$ and $Z_L=\otimes_{i=1}^{n} Z_i$, as in the Steane $[[7,1,3]]$ code.  Our notation in this paper extends the usual one slightly, since we wish also to detect up to $s$ syndrome measurement errors - hence we refer to $[[n,k,d,s]]$ codes.  Standard quantum error correction without repetitions has $s=0$.  

In DBR for a CSS (CSS-DBR) code we measure $m=C/2$ stabilizer operators of the form 
$S = X_{i_1} X_{i_2} \cdots X_{i_w}$
and $m=C/2$ stabilizer operators of the form
$S = Z_{i_1} Z_{i_2} \cdots Z_{i_w}$. Each stabilizer has weight $w$. Here $\{i_1, i_2,..., i_w\}$ are chosen from the set $\{1,2,\dots,n\}$, and any given $S$ is completely defined by this choice.   Thus for CSS codes X and Z errors are handled separately, so we simply use 2 copies of a single design, and this gives a particularly economical and effective DBR protocol. The measurement result of the $Z$-type stabilizers is different for any bit flip error.  Tabulating these possible syndrome results yields an $m \times (n+1) $ matrix $E$ whose entries are $\pm 1$.  The entry $E_{ij}$ is the result of correctly extracting the value of $S_i$ for a bit flip on the $j$-th physical qubit.  At a minimum, the rows of $E$ must all be distinct in order to diagnose an error uniquely.  We wish to go beyond this.  \textit{Fault tolerance in DBR is achieved not by repetition but by choosing our measurements so that the results differ by as much as possible.} This motivates the use of 2-designs for our choice of measurements.

A 2-design is a family of subsets of a larger set.  This family must fulfill certain conditions.  For present purposes there is a set of $n$ qubits and the indices on each $Z$-type stabilizer to be measured defines a subset of the qubit indices.  Thus the family of subsets is determined by the choice of $Z$-type stabilizers.  There are $m$ subsets.  The conditions for this choice to be a 2-design are: (1) each subset has the same size $w$; (2) every index must appear in exactly $\rho$ subsets; (3) every pair of indices appears in exactly $\lambda$ subsets.  The parameters are not all independent.  They satisfy the basic 2-design relations $mw=n \rho$ and $\lambda (n-1) = \rho (w-1)$.  These equations are proved using counting arguments.  

 Let us define the Hamming-like distance $D$ between any two $Z$-type stabilizers $S_{i}$ and $S_{i'}$ as $D(S_{i},S_{i'}) = \Sigma_{j=1}^{r} |E_{ij}-E_{i'j}|$, where $E_{ij}$ is the result when the $j$th qubit has flipped.  Then if the choice of the $S_i$ is a 2-design we have that $D(S,S') =2(\rho-\lambda)$ for all $S$ and $S'$.  This may be shown by arguments similar to those in \cite{pless}.  This the key feature of 2-designs for error correction purposes: it enables us to systematically maximize $\rho - \lambda $. 

To illustrate the definition we give the example of a simple 2-design known as the order-2 biplane.  It applies to a system of $n=7$ qubits.  The stabilizers are $S_1 = Z_1 Z_5 Z_6 Z_7$, $S_2 = Z_2 Z_4 Z_6 Z_7$, $S_3 = Z_3 Z_4 Z_5 Z_7$, $S_4 = Z_1 Z_2 Z_4 Z_5$, $S_5 = Z_1 Z_3 Z_4 Z_6$, $S_6 = Z_2 Z_3 Z_5 Z_6$, $S_7 = Z_1 Z_2 Z_3 Z_7$.  It is not hard to verify that this choice satisfies the constraints for a 2-design with $w=4$, $m=7$, $\rho = 4$ and $\lambda = 2$.  We will use this 2-design below. 

The fundamental criterion for the choice of stabilizers is the minimization of the failure rate $F$ and the cost $C$.  We have seen that the natural arena for this is the 2-design.  However, not all 2-designs can be used in conjunction with stabilizer quantum error correction.  There is one constraint for all DBR schemes based on 2-designs.

\textbf{Constraint 1.}
The definition of a stabilizer requires that it must commute with the logical operators, \textit{i.e.}, $[S, \otimes_{i=1}^{n}X_i] = [S, \otimes_{i=1}^{n}Z_i]=0$ for all $S$.  A short calculation shows that this is true if and only if $w$ is even.  (See supplementary material.)

\textbf{Constraint 2.}
(CSS-DBR only.) The use of two copies of a design generates an additional constraint. The stabilizer group is commutative.  All $X$-type stabilizers trivially commute with each other; the same is true for the $Z$-type stabilizers.  An $X$-type stabilizer commutes with a $Z$-type stabilizer if and only if the intersection of the set of indices of the $X$-type stabilizer with the set of indices of the $Z$-type stabilizer has even cardinality.  This happens for all such pairs of stabilizers if and only if $\lambda$ is even. 

These two constraints rule out a majority of 2-designs for CSS-DBR.  The order-2 biplane with $w=4$ and $\lambda=2$ is allowed, but the natural successor is the order-3 biplane with $w=5$: it cannot serve as the basis for an DBR scheme.

\section{Results}

\subsection{5-qubit code}

The [[5,1,3]] perfect code is not a CSS code, so a single 2-design is used for all errors.  We take the stabilizer group generated by $S_1 = X_1 Z_2 Z_3 X_4, S_2 = X_2 Z_3 Z_4 X_5, S_3 = X_1 X_3 Z_4 Z_5,$ and $S_4 = Z_1 X_2 X_4 Z_5$, which is then a [[5,1,3,1]] code. The full stabilizer group is an instance of the class of designs generated from Hadamard matrices, appropriately called Hadamard designs. In particular, the design here is the complement to the $n = 3$ Hadamard design \cite{Dinitz1992}.  Once the stabilizer set is chosen we can compute the failure rate $F_{DBR}$ as a function of $p_m$ and $p_q$.  DBR stands for "design-based redundancy".  We also compute the failure rate $F_{MR}$ for a scheme in which minimal redundancy (MR) is employed: only one additional stabilizer is measured, namely $ S_1 S_2 S_3 S_4$.  This operator is the product of the minimal set of generators.  With this choice the resulting protocol is somewhat analogous to a single parity check in classical error correction.  Finally we compute the failure rate $F_{QEC}$ for fault-tolerant QEC.  Analytical results for all three rates are given in the supplementary material.  Choosing $F_{QEC}$ as a basline, we plot the relative failure rates in Fig. 1 for a range of error probabilities relevant to near-term quantum information processing. Note that MR has an advantage over QEC for $p_m > XXX p_q$. Both are fault-tolerant with respect to $p_m$ in that $F_{QEC}. F_{MR} \sim p_m^2$. However, DBR is superior to both QEC and MR for all $p_m, p_q$ in the appropriate regime.  This can be traced back to the fact that $F_{DBR} \sim p_m^4$. Note that MR is cheaper than DBR in terms of number of measurements, so it may be preferable in intermediate regimes of $p_m$.            

\begin{figure}[h]
	\centering
	\includegraphics[width=\linewidth]{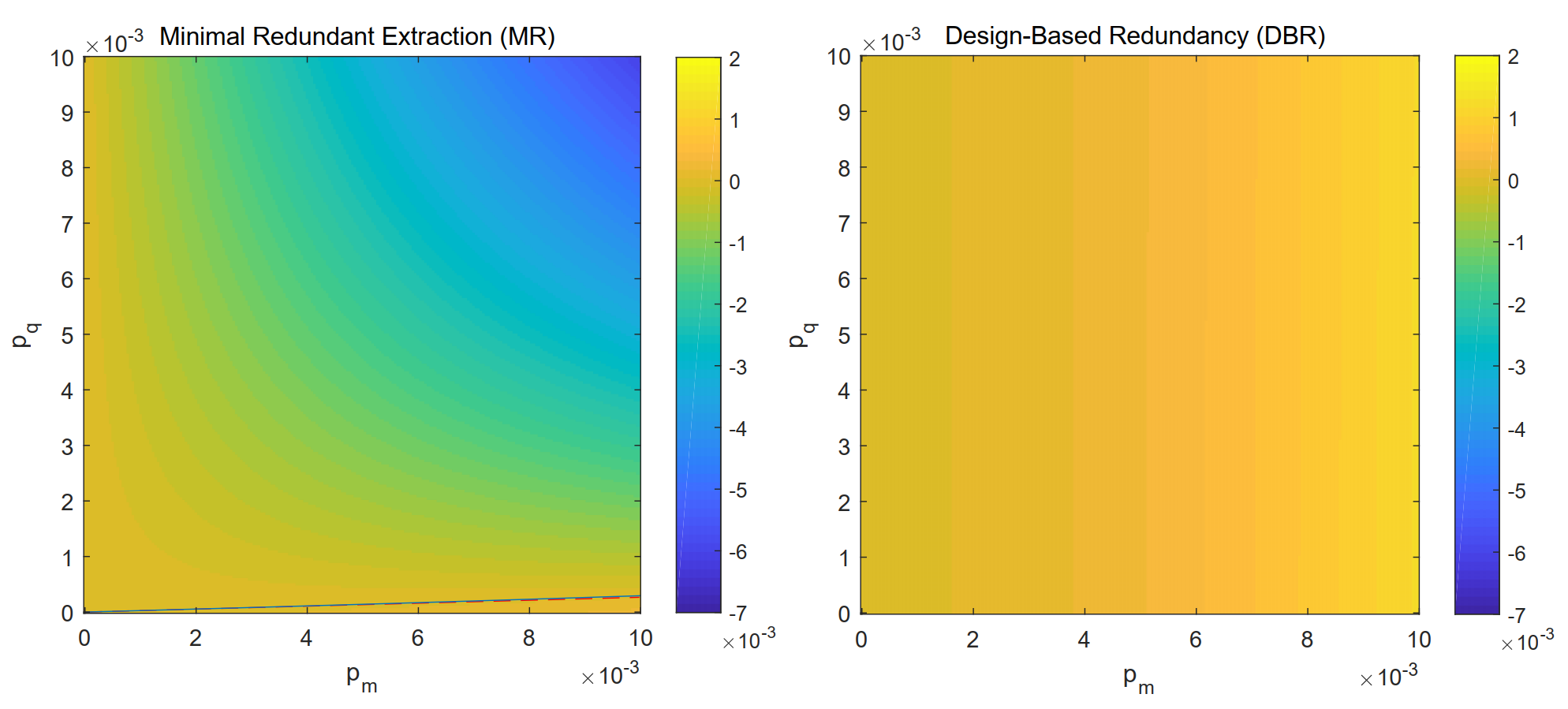}
	\caption{$F_{QEC} - F_{MR}$ and $F_{QEC} - F_{DBR}$ for the [[5,1,3]] Perfect code. The left plot shows a curve dividing the parameter space into configurations where QEC outperforms MR, whereas at all physical error rates DBR beats QEC. This comes at the cost of more stabilizer measurements. }
\label{fig:5qubit}
\end{figure}

\subsection{Steane code}

In our notation, the Steane code with the minimal set of measurements is a $[[7,1,3,0]]$ code.  When syndrome measurements errors occur, it is not fault-tolerant.  3-fold repetition of its measurement sequence with majority rules gives a $[[7,1,3,1]]$ code.  The order-2 biplane DBR procedure utilizes 14 stabilizers: those given in the previous section and another 7 with $Z_i \rightarrow X_i$.  It is a fault tolerant $[[7,1,3,1]]$ code. 
\begin{table}[h]
\begin{tabular} {ccc}
\toprule
Protocol & Failure Rate & Cost \\
\midrule
Minimal QEC & $3p_m - 3p_m^2 + 21p_q^2$  &  $3$ \\
Fault-tolerant QEC  & $9 p_m^2 + 21 p_q^2$  &  $6 + 3 p_m$  \\

MR with $S_7$ & $6p_m^2 + 21p_q^2 + 28 p_m p_q$ & $4$ \\

DBR  &  $21 p_q^2$  &  7\\
\bottomrule
\end{tabular}
	\caption {Failure rate and cost for 4 error-correction protocols 
	for the Steane [[7,1,3]] and [[7,1,3,1]] codes.}
\end{table}
We again give the comparison of failure rates for 3 protocols: the repeated Steane code, denoted QEC; an MR code with two additional stabilizers, one of the X type and one of the Z type, in both cases a product of the usual generators; and the MR approach measuring the full stabilizer group. This final set of stabilizers corresponds to 2 copies of the order-2 biplane.  The results demonstrate that DBR has a distinct advantage over the other 2 protocols.  It is even more dramatic than in the 5-qubit case.  This combination of DBR with a CSS code is particularly effective in repairing measurement errors.  Again, the MR approach may also be useful in practical cases - everything depends on the cost and reliability of measurements in a specific implementation.

\begin{figure}[h]
	\centering
	\includegraphics[width=\linewidth]{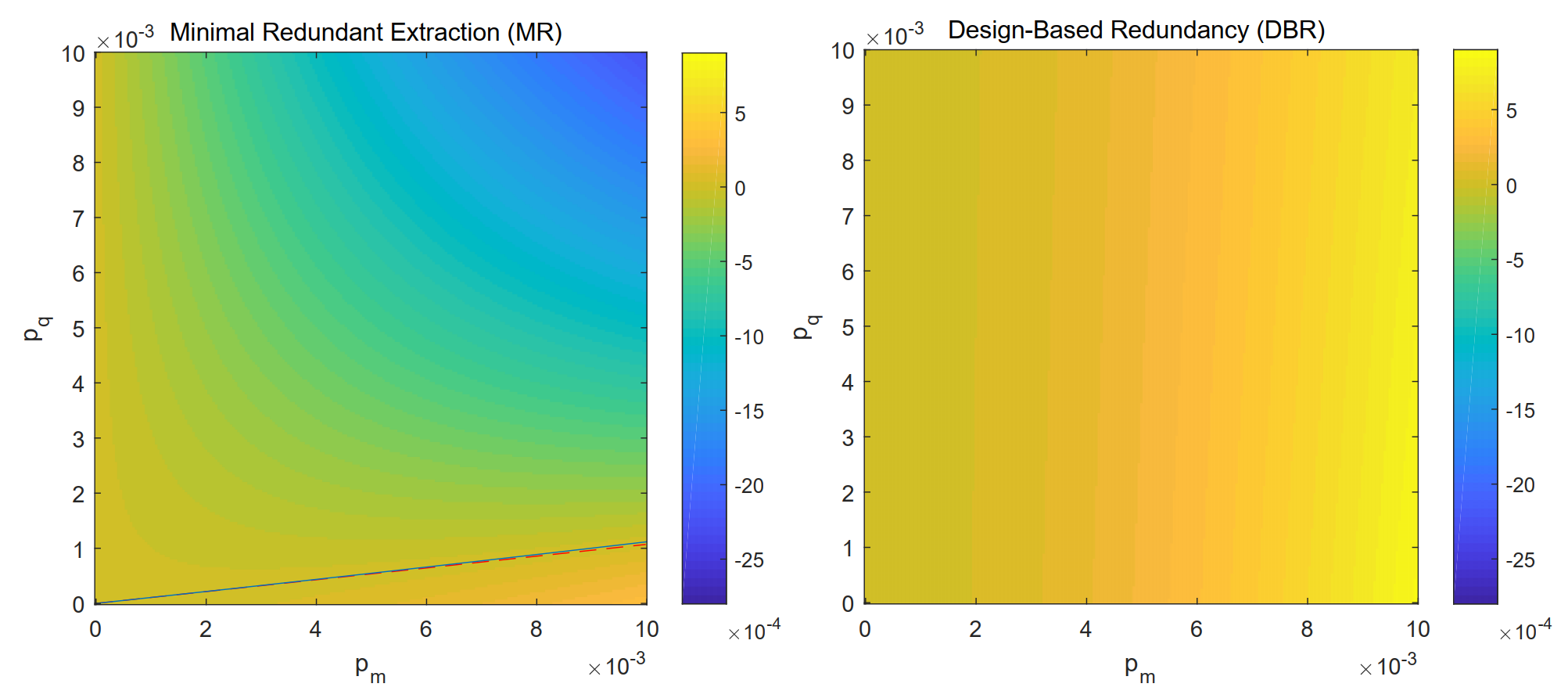}
	\caption{$F_{QEC} - F_{MR}$ and $F_{QEC} - F_{DBR}$ for the [[7,1,3]] Steane code. The left plot shows a curve dividing the parameter space into configurations where QEC outperforms MR, whereas at all physical error rates DBR beats QEC. This comes at the cost of more stabilizer measurements.}
\label{fig:Steane}
\end{figure}

\section{Conclusion}

In this paper we have focused on 2-designs that appear naturally in straightforward modifications of well-known codes.  But a longer-term goal would be to use known 2-designs to improve syndrome extraction, thus improving quantum error-correction in cases where errors in the measurement process are important. This not straightforward, since there is no general classification theorem for 2-designs, though a large number of special cases and some infinite families are known \cite{Rudolf1985}.  This is a promising line of research.

Once a quantum processor is characterized and $p_q, p_m$ are known, we can look at the failure rates to determine which approach to take. The next consideration is cost. In a Shor-style extraction protocol each stabilizer measurement requires a number of ancilla qubits one greater than the weight of the operator itself \cite{Shor1996}. In fault-tolerant QEC ancilla qubits can be reused on subsequent cycles of syndrome extraction. Thus, even though the number of QEC measurements may exceed that of DBR for a given code, the size of the quantum register required may in fact be larger for DBR. Recent ideas for using fewer ancilla qubits could be used to bring down the potential cost disparity between QEC and DBR \cite{Chao2018}. This comes at the price of adding more gates, so an appraisal of which is the most fruitful approach involves a balancing act between number of qubits and circuit depth. 

In the near term, we may expect the development of machines with 50-100 fairly noisy qubits and gates (the "NISQ" era).  One important goal for this era is the development of a single fault-tolerant error-corrected logical qubit.  The correction process involves multiple gates while the data qubits will often have reasonably long coherence times.  In this situation, probably the most common one, we expect $p_m > p_q$.  This means that the strategies outlined in this paper will be very relevant for this important goal.    

\begin{acknowledgments}
	This research was sponsored in part by the Army Research Office (ARO) under Grant Numbers W911NF-17-1-0274.The views and conclusions contained in this document are those of the authors and should not be interpreted as representing the official policies, either expressed or implied, of the Army Research Office (ARO), or the U.S. Government. The U.S. Government is authorized to reproduce and distribute reprints for Government purposes notwithstanding any copyright notation herein. 
\end{acknowledgments}
\setcounter{secnumdepth}{0} 
\section{Author Contributions}
H.S. performed the numerical simulations and V.N.P and R.J. worked on analytic calculations, analyzed the results and prepared the manuscript. D.C. provided insight on the structure of stabilizer codes as viewed through the lens of 2-designs and E.B. supported the work with a significant body of knowledge relating to classical coding. The entire work was carried out under the supervision of R.J.

\setcounter{secnumdepth}{0} 
\section{Competing Interests}
The authors declare no competing interests.

\bibliographystyle{unsrt}
\bibliography{bibliography}

\end{document}


\title{Supplemental Material for 2-designs and Redundant Syndrome Extraction for Quantum Error Correction }
\author{Vickram N. Premakumar}
\address{Physics Department, University of Wisconsin-Madison, 1150 Univ. Ave., Madison, WI, USA}
\author{Hele Sha}
\address{Physics Department, University of Wisconsin-Madison, 1150 Univ. Ave., Madison, WI, USA}
\author{Daniel Crow}
\address{Physics Department, University of Wisconsin-Madison, 1150 Univ. Ave., Madison, WI, USA}
\author{Eric Bach}
\address{Computer Science Department, University of Wisconsin-Madison, Madison, WI, USA}
\author{Robert Joynt}
\address{Physics Department, University of Wisconsin-Madison, 1150 Univ. Ave., Madison, WI, USA}
\address{
	Kavli Institute for Theoretical Sciences, University of Chinese Academy of Sciences, Beijing 100190, China}
\date{{\normalsize \today}}

\maketitle

\section{Constraints on BIBD Parameters for QEC}

\subsection{Proof of constraint 1}
We must show that all the measured stabilizers $S$ commute with all the logical operators $L$ if and only if $w$ is even.
An example of such a commutator is 
\begin{equation*}
[S,L] = [ Z_{i_1} Z_{i_2} ... Z_{i_w}, X_1 X_2... X_n ],
\end{equation*}
where $ \{i_1,i_2,...i_w \} \subset  \{1,2,...,n\}$.  We then define the set of indices $\{k_1,k_2...k_{(n-w)} \}$ such that 
$ \{i_1,i_2,...i_w \} \cup \{j_1,j_2...j_{(n-w)} \} = \{1,2...n\}$
This can be simplified as follows: 
\begin{eqnarray*}
	[S,L] 
	& = & [Z_{i_1} Z_{i_2} ... Z_{i_w}, X_1 X_2... X_n] \\
	& = & X_{k_1} X_{k_2}... X_{k_{(n-w)}} [Z_{i_1} Z_{i_2} ... Z_{i_w}, X_{i_1} X_{i_2}... X_{i_w}] \\
	& = &  X_{k_1} X_{k_2}... X_{k_{(n-w)}}  \times \\
	&  &  (Z_{i_1} Z_{i_2} ... Z_{i_w}X_{i_1} X_{i_2}... X_{i_w} -  X_{i_1} X_{i_2}... X_{i_w} Z_{i_1} Z_{i_2} ... Z_{i_w}) \\
	& = &  X_{k_1} X_{k_2}... X_{k_{(n-w)}} \times \\
	& &  (Z_{i_1} X_{i_1} Z_{i_2} X_{i_2} ... Z_{i_w}X_{i_w} -  X_{i_1} Z_{i_1} X_{i_2} Z_{i_2}... X_{i_w} Z_{i_w}) \\
	& = &  X_{k_1} X_{k_2}... X_{k_{(n-w)}} \times \\
	& &  ((-1)^w X_{i_1} Z_{i_1} X_{i_2} Z_{i_2} ... X_{i_w}Z_{i_w} -  X_{i_1} Z_{i_1} X_{i_2} Z_{i_2}... X_{i_w} Z_{i_w}) \\
	& = &  X_{k_1} X_{k_2}... X_{k_{(n-w)}} [ (-1)^w -1] X_{i_1} Z_{i_1} X_{i_2} Z_{i_2} ... X_{i_w}Z_{i_w},
\end{eqnarray*}
which is zero if and only if $w$ is even.  This proof clearly holds for all stabilizer/logical operator pairs.

\subsection{Proof of constraint 2}
For the CSS-DBR codes, the same design is used twice, once for the X-stabilizers and once for the Z-stabilizers.  The full group must be abelian, so one must check that the X-stabilizers commute with the Z-stabilizers. The relevant commutators have the form 
\begin{equation*}
C =  [ X_{i_1} X_{i_2} ... X_{i_w} , Z_{j_1} Z_{j_2} ... Z_{j_w} ] 
\end{equation*}
Here $ \{i_1,i_2,...i_w \} \subset  \{1,2,...,n\}$. and $ \{j_1,j_2,...j_w \} \subset  \{1,2,...,n\}$. Inspection of the proof for constraint 1 then indicates that if we define 
\begin{equation*}
x = | \{i_1,i_2,...i_w \} \cap \{j_1,j_2,...j_w \} |,
\end{equation*}
then $C=0$ if and only if $x$ is even.  This is true for all choices of the index sets if $\lambda$ is even.   

\section{Failure Rates}

In the main text we showed the failure rates of traditional QEC compared with two proposed extensions by plotting the difference as a function of both physical and measurement error rates. To emphasize the region of interest for near-threshold quantum computers, we restricted $p_q, p_m$ to a small range close to zero. For completeness, we include the full comparison over the entire feasible parameter range. When the curve where $F_{QEC} - F_{MR/DBR} = 0$ does not pass through the origin, this reflects a difference in the power of the leading order $p_q$ or $p_m$ between the two rates. That is to say, one method can tolerate more errors of a certain type than the other when this is the case.

\begin{figure}
	\centering
	\includegraphics[width=\linewidth]{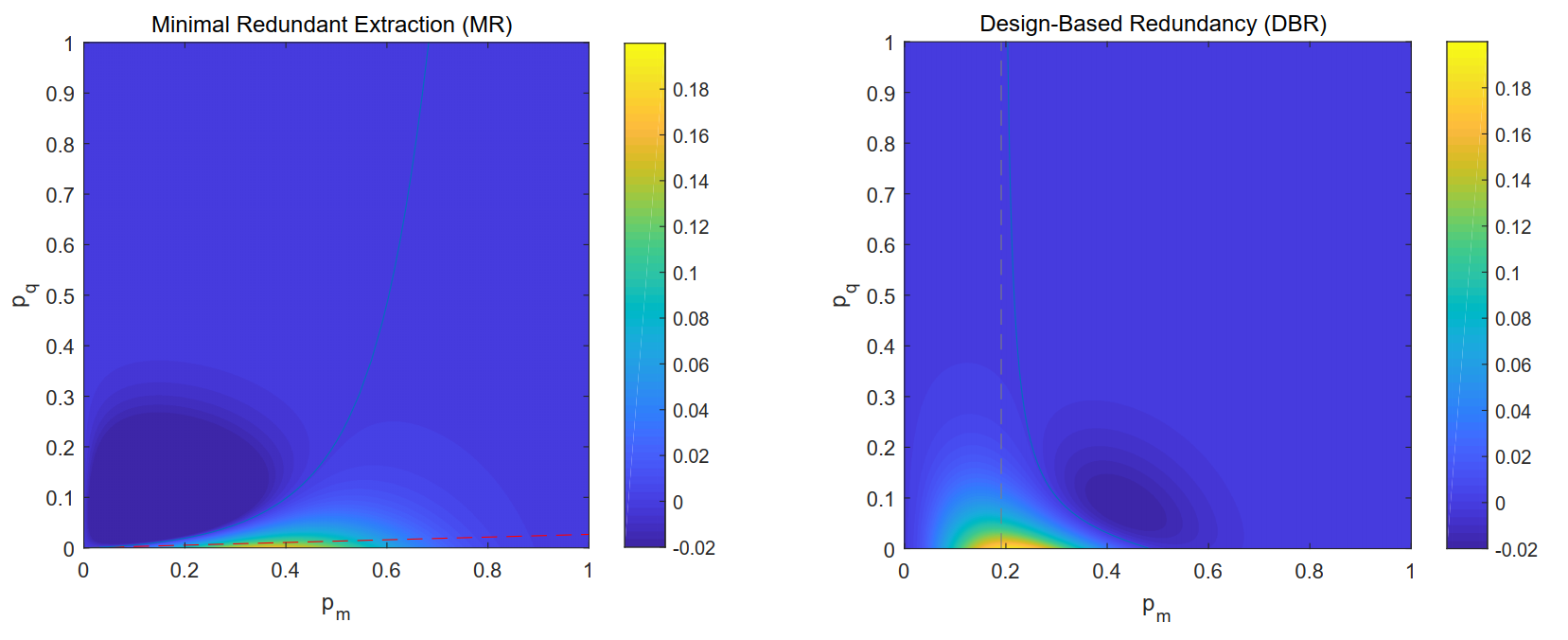}
	\caption{$F_{QEC} - F_{MR}$ and $F_{QEC} - F_{DBR}$ for the [[5,1,3]] Perfect code. This is a zoomed-out version of Fig. 1 in the main text.  Here we  show the results for all possible error rates.}
	\label{fig:5qubitperfectcodefullrange}
\end{figure}

\begin{figure}
	\centering
	\includegraphics[width=\linewidth]{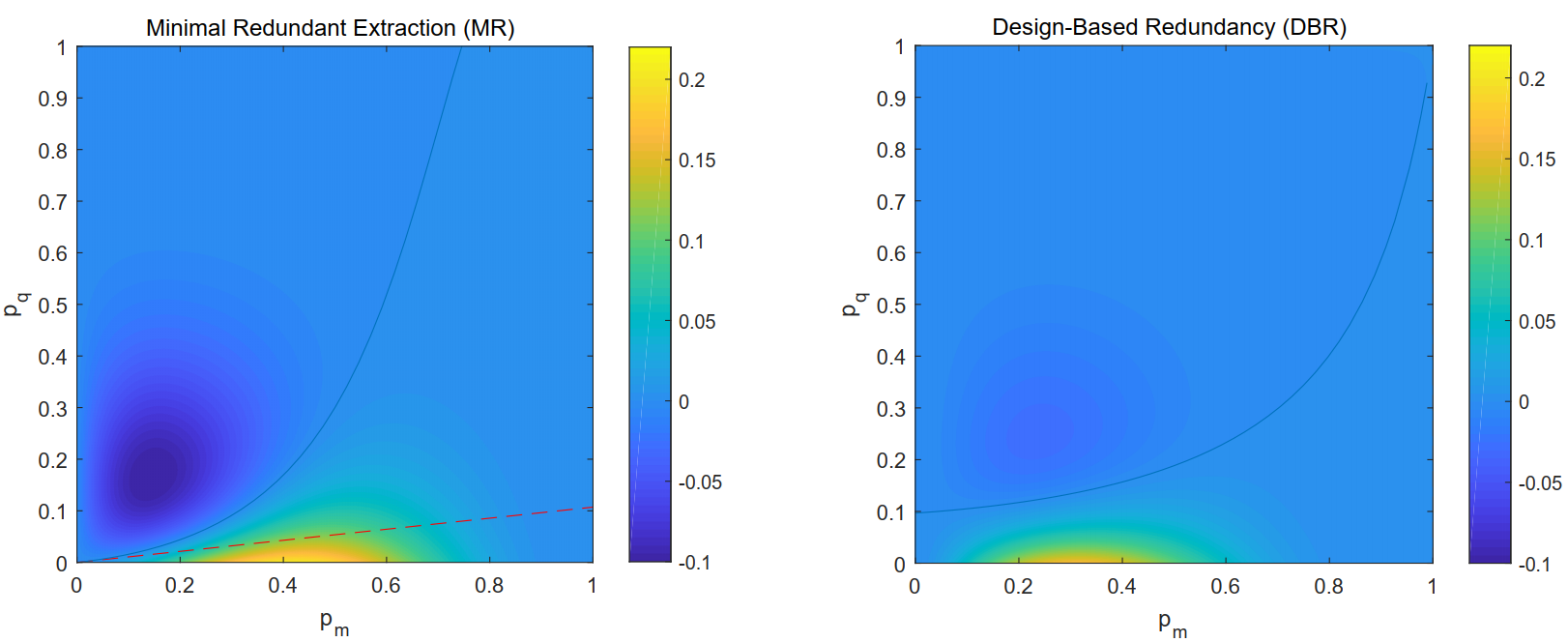}
	\caption{$F_{QEC} - F_{MR}$ and $F_{QEC} - F_{DBR}$ for the [[7,1,3]] Steane code. This is a zoomed-out version of Fig. 2 in the main text.  Here we show the results for all possible error rates.}
	\label{fig:steanecodefullrange}
\end{figure}

The main text has truncated analytic forms for the failure rates as a function of $p_q, p_m$, Here are the expressions to order $p^8$.  The full expressions to all orders are available on request to the authors.

For 5-Qubit Perfect Code:

\begin{align*} 
F_{QEC} &= 12p_{m}^{2} - 8p_{m}^{3} - 54p_{m}^{4} + 72p_{m}^{5} + 84p_{m}^{6} - 216p_{m}^{7} + 63p_{m}^{8} + 105p_{q}^{2} - 910p_{q}^{3} + 4095p_{q}^{4} - 12012p_{q}^{5} + 25025p_{q}^{6} - 38610p_{q}^{7} \\ & + 45045p_{q}^{8} - 1260p_{m}^{2}p_{q}^{2} + 10920p_{m}^{2}p_{q}^{3} + 840p_{m}^{3}p_{q}^{2} - 49140p_{m}^{2}p_{q}^{4} - 7280p_{m}^{3}p_{q}^{3} + 5670p_{m}^{4}p_{q}^{2} + 144144p_{m}^{2}p_{q}^{5} + 32760p_{m}^{3}p_{q}^{4} \\ & - 49140p_{m}^{4}p_{q}^{3} - 7560p_{m}^{5}p_{q}^{2} - 300300p_{m}^{2}p_{q}^{6} - 96096p_{m}^{3}p_{q}^{5} + 221130p_{m}^{4}p_{q}^{4} + 65520p_{m}^{5}p_{q}^{3} - 8820p_{m}^{6}p_{q}^{2} 
\end{align*}

\begin{align*}
 F_{DBR} &=  105p_{q}^{2} - 910p_{q}^{3} + 4095p_{q}^{4} + 20475p_{m}^{4}p_{q} - 225225p_{m}^{5}p_{q} + 1126125p_{m}^{6}p_{q} - 3378375p_{m}^{7}p_{q} + 3003p_{m}^{5} - 25025p_{m}^{6} + 96525p_{m}^{7} \\ & - 225225p_{m}^{8}  - 12012p_{q}^{5} + 25025p_{q}^{6} - 38610p_{q}^{7} + 45045p_{q}^{8} - 286650p_{m}^{4}p_{q}^{2} + 1863225p_{m}^{4}p_{q}^{3} + 2837835p_{m}^{5}p_{q}^{2} - 7452900p_{m}^{4}p_{q}^{4}  \\ & -  17762745p_{m}^{5}p_{q}^{3} - 13138125p_{m}^{6}p_{q}^{2} 
\end{align*}

\begin{align*}
 F_{MR} &= 75p_{m}p_{q} - 1050p_{m}p_{q}^{2} - 300p_{m}^{2}p_{q} + 6825p_{m}p_{q}^{3} + 450p_{m}^{3}p_{q} - 27300p_{m}p_{q}^{4} - 300p_{m}^{4}p_{q} + 75075p_{m}p_{q}^{5} + 75p_{m}^{5}p_{q} - 150150p_{m}p_{q}^{6} \\ & + 225225p_{m}p_{q}^{7} + 10p_{m}^{2} - 20p_{m}^{3} + 15p_{m}^{4} - 4p_{m}^{5} + 105p_{q}^{2} - 910p_{q}^{3} + 4095p_{q}^{4} - 12012p_{q}^{5} + 25025p_{q}^{6}  - 38610p_{q}^{7} + 45045p_{q}^{8} \\ & + 3150p_{m}^{2}p_{q}^{2} - 18200p_{m}^{2}p_{q}^{3} - 4200p_{m}^{3}p_{q}^{2} + 68250p_{m}^{2}p_{q}^{4} + 22750p_{m}^{3}p_{q}^{3} + 2625p_{m}^{4}p_{q}^{2} - 180180p_{m}^{2}p_{q}^{5} - 81900p_{m}^{3}p_{q}^{4} \\ & - 13650p_{m}^{4}p_{q}^{3} - 630p_{m}^{5}p_{q}^{2} + 350350p_{m}^{2}p_{q}^{6} + 210210p_{m}^{3}p_{q}^{5} + 47775p_{m}^{4}p_{q}^{4} + 3185p_{m}^{5}p_{q}^{3} 
\end{align*}

For Steane Code:

\begin{align*} F_{QEC} &= 9p_{m}^{2} - 6p_{m}^{3} - 27p_{m}^{4} + 36p_{m}^{5} + 15p_{m}^{6} - 54p_{m}^{7} + 36p_{m}^{8} + 21p_{q}^{2} - 70p_{q}^{3} + 105p_{q}^{4} - 84p_{q}^{5} + 35p_{q}^{6} - 6p_{q}^{7} - 189p_{m}^{2}p_{q}^{2} \\ &+ 630p_{m}^{2}p_{q}^{3} + 126p_{m}^{3}p_{q}^{2} - 945p_{m}^{2}p_{q}^{4} - 420p_{m}^{3}p_{q}^{3} + 567p_{m}^{4}p_{q}^{2} + 756p_{m}^{2}p_{q}^{5} + 630p_{m}^{3}p_{q}^{4} - 1890p_{m}^{4}p_{q}^{3} - 756p_{m}^{5}p_{q}^{2} - 315p_{m}^{2}p_{q}^{6} \\ &- 504p_{m}^{3}p_{q}^{5} + 2835p_{m}^{4}p_{q}^{4} + 2520p_{m}^{5}p_{q}^{3} - 315p_{m}^{6}p_{q}^{2}
\end{align*}

\begin{align*} F_{DBR} &= 147p_{m}^{2}p_{q} - 735p_{m}^{3}p_{q} + 1470p_{m}^{4}p_{q} - 1470p_{m}^{5}p_{q} + 735p_{m}^{6}p_{q} - 147p_{m}^{7}p_{q} + 35p_{m}^{3} - 105p_{m}^{4} + 126p_{m}^{5} - 70p_{m}^{6} + 15p_{m}^{7} \\ & + 21p_{q}^{2} - 70p_{q}^{3} + 105p_{q}^{4} - 84p_{q}^{5} + 35p_{q}^{6} - 6p_{q}^{7} - 882p_{m}^{2}p_{q}^{2} + 2205p_{m}^{2}p_{q}^{3} + 3675p_{m}^{3}p_{q}^{2} - 2940p_{m}^{2}p_{q}^{4} - 8575p_{m}^{3}p_{q}^{3} - 6615p_{m}^{4}p_{q}^{2} \\ & + 2205p_{m}^{2}p_{q}^{5} + 11025p_{m}^{3}p_{q}^{4} + 14700p_{m}^{4}p_{q}^{3} + 6174p_{m}^{5}p_{q}^{2} - 882p_{m}^{2}p_{q}^{6} - 8085p_{m}^{3}p_{q}^{5} - 18375p_{m}^{4}p_{q}^{4} - 13230p_{m}^{5}p_{q}^{3} - 2940p_{m}^{6}p_{q}^{2} 
\end{align*}

\begin{align*} F_{MR} &= 28p_{m}p_{q} - 168p_{m}p_{q}^{2} - 84p_{m}^{2}p_{q} + 420p_{m}p_{q}^{3} + 84p_{m}^{3}p_{q} - 560p_{m}p_{q}^{4} - 28p_{m}^{4}p_{q} + 420p_{m}p_{q}^{5} - 168p_{m}p_{q}^{6} + 28p_{m}p_{q}^{7} + 6p_{m}^{2} \\ & - 8p_{m}^{3} + 3p_{m}^{4} + 21p_{q}^{2} - 70p_{q}^{3} + 105p_{q}^{4} - 84p_{q}^{5} + 35p_{q}^{6} - 6p_{q}^{7} + 378p_{m}^{2}p_{q}^{2} - 840p_{m}^{2}p_{q}^{3} - 336p_{m}^{3}p_{q}^{2} + 1050p_{m}^{2}p_{q}^{4} + 700p_{m}^{3}p_{q}^{3} \\ &+ 105p_{m}^{4}p_{q}^{2} - 756p_{m}^{2}p_{q}^{5} - 840p_{m}^{3}p_{q}^{4} - 210p_{m}^{4}p_{q}^{3} + 294p_{m}^{2}p_{q}^{6} + 588p_{m}^{3}p_{q}^{5} + 245p_{m}^{4}p_{q}^{4} 
\end{align*}

For Bitflip Code:

\begin{align*} F_{QEC} &= 6p_{m}^{2} - 4p_{m}^{3} - 9p_{m}^{4} + 12p_{m}^{5} - 4p_{m}^{6} + 3p_{q}^{2} - 2p_{q}^{3} - 18p_{m}^{2}p_{q}^{2} + 12p_{m}^{2}p_{q}^{3} \\ & + 12p_{m}^{3}p_{q}^{2} - 8p_{m}^{3}p_{q}^{3} + 27p_{m}^{4}p_{q}^{2} - 18p_{m}^{4}p_{q}^{3} - 36p_{m}^{5}p_{q}^{2} + 24p_{m}^{5}p_{q}^{3} + 12p_{m}^{6}p_{q}^{2} 
\end{align*}

\[ F_{DBR} = 9p_{m}p_{q} - 18p_{m}p_{q}^{2} - 18p_{m}^{2}p_{q} + 9p_{m}p_{q}^{3} + 9p_{m}^{3}p_{q} + 3p_{m}^{2} - 2p_{m}^{3} + 3p_{q}^{2} - 2p_{q}^{3} + 27p_{m}^{2}p_{q}^{2} - 12p_{m}^{2}p_{q}^{3} - 12p_{m}^{3}p_{q}^{2} + 5p_{m}^{3}p_{q}^{3} \]